\documentclass{mem}
\usepackage{natbib}\usepackage{txfonts}\usepackage{balance}
\usepackage{graphicx}
\usepackage{txfonts}
\usepackage[a4paper]{hyperref}
\idline{79}{3}
\begin{document}
\def\teff{$T\rm_{eff }$}
\def\kms{$\mathrm {km s}^{-1}$}

\title{The Peculiar Horizontal Branch Morphology of the Galactic Globular Clusters NGC~6388 and NGC~6441}

   \subtitle{}

\author{
G. \,Busso%\inst{1} 
}

  \offprints{G. Busso}

\institute{
Istituto Nazionale di Astrofisica --
Osservatorio Astronomico di Teramo, Via Mentore Maggini s.n.c,
I-64100 Teramo, Italy
%\and
%Forest  University, Department of Astronomy,
%25 Long Street, 255255,
%Somewhere, Elsewhere 
\email{busso@oa-teramo.inaf.it}
}

\authorrunning{Busso,G. et al.}

\titlerunning{The peculiar HB morphology of NGC~6388 and NGC~6441}

\abstract{I present multiband optical and UV Hubble Space
Telescope photometry of the two Galactic globular clusters NGC~6388
and NGC~6441, in order to investigate the nature
of the physical mechanism(s) responsible for the
existence of an extended blue tail and of a slope in the horizontal branch.
Further evidence that the horizontal branch tilt cannot be
interpreted as an effect of differential reddening is provided, while I show
that a possible solution of the puzzle is to assume that a
small fraction of the stellar population in the two clusters is strongly helium enriched
(Y$\sim$0.40 in NGC~6388 and Y$\sim$0.35 in NGC~6441).
}

\maketitle{}

\section{Introduction}

This work is part of an HST project devoted to study some globular clusters
(namely NGC~6388, NGC~6441, NGC~6273 and NGC~5986), that show an extended horizontal branch (HB). In this paper I present the results for NGC~6388 and NGC~6441, while the results for the other two clusters will be published in a later paper.

The data are a merge of optical and UV observations, both collected with the WFPC2@HST. The optical data (in the filters F439W and F555W) are part of the previous Galactic Globular Clusters (GGCs) survey obtained from Piotto et al.~2002.  
The UV data represent an extension of the
original F439W-F555W vs. F555W CMDs published by Rich et al.~(1997) and Piotto et al.~(2002). Because of the strong bolometric correction and temperature
insensitivity, the optical data are not as useful in the study of the
hottest HB stars, which can be better studied instead with the UV observations. 

NGC~6388 and NGC~6441 are peculiar for several reasons:

\begin{itemize}
\item both clusters show not only the red HB, characteristic of metal-rich GCs as 47~Tuc ([Fe/H]$=-0.76$,  Harris 1996) but also a quite extended blue HB. 
\item the mean HB  brightness at the top of the blue HB tail is roughly 0.5 magnitudes brighter in the F555W band than the red HB portion, which appears significantly sloped as well. 
  This slope, as pointed out in Sweigart \& Catelan 1998 is not present in the theoretical models, and is difficult to explain (see also the discussion in Raimondo et al.~2002). 
\end{itemize}

In an attempt to interpret the occurrence of the blue tail along the
HB as well as of its tilted morphology, many non-canonical scenarios
and/or observational effects have been suggested, as for example  the initial He abundance, a spread in metallicity, differential reddening, rotation, etc. (see Catelan~2007 for a review).  

%%%%%%% TABLE 1 %%%%%%%%%
\begin{table}
\caption{Fundamental parameters of NGC~6388 and NGC~6441\protect\footnotemark{}.}
\begin{center}
\begin{tabular}{|c|c|c|}
\hline 
   & NGC~ 6388 & NGC~6441 \\
\hline 
$l$  & 345.56 & -353.53 \\
\hline 
$b$  & -6.74 & -5.01 \\
\hline 
$d_{\odot}$ & 10.0 & 11.7 \\
\hline 
[Fe/H] & -0.60 & -0.53 \\
\hline 
 $E(B-V)$ & 0.37 & 0.47 \\
\hline 
\end{tabular}
\end{center}
\end{table}

 %footnote in the caption of Tab.1
\footnotetext{l and b are respectively galactic longitude and latitude (in degrees), $d_{\odot}$ is the
distance from the Sun (in Kpc), [Fe/H] is the metallicity, and
$E(B-V)$ is the reddening (from the latest version of the Harris 1996
catalog).}
%%%%%%%%%%%%%%%%%%%%%%%%
The main parameters of NGC~6388 and NGC~6441, as given
by  Harris (1996, in the 2003 revision), are summarized in Table 1.
The data come from
HST/WFPC2 observations; in all cases, the PC camera was centered on the
cluster center. The photometric reduction was carried out using the {\small
DAOPHOT}II/{\small ALLFRAME} package (Stetson 1987, 1994). For more details about the data reduction see Busso et al.~2007.

%\section{The anomalous HBs of NGC~6388 and NGC~6441}

\section{The color-magnitude diagrams}
\begin{figure*}
\begin{center}
\includegraphics[width=10cm,height=9cm]{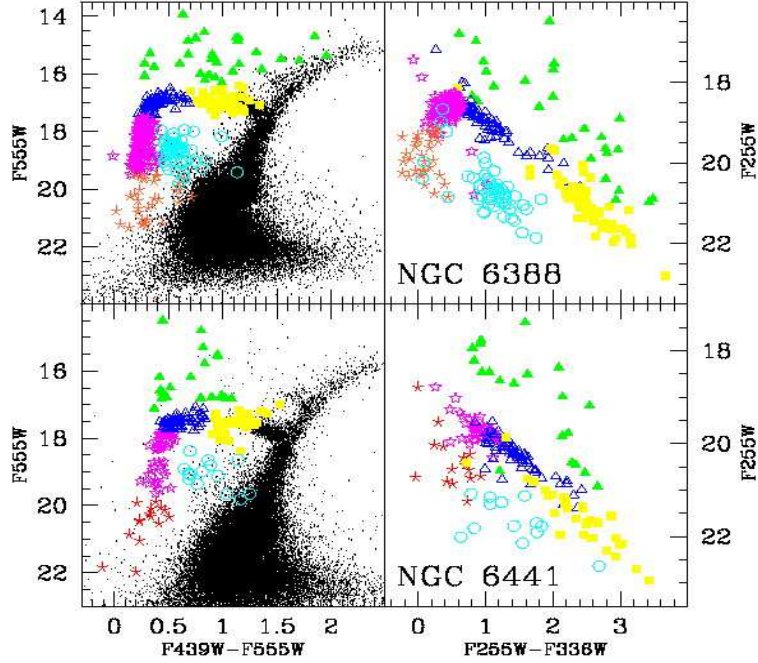}
\end{center}
\caption{\footnotesize
Optical and UV CMDs for NGC~6388 (upper panels) and NGC~6441
(lower panels).  The different symbols refer to the same
group of stars appearing both in the optical and UV CMD: filled squares indicate the red HB, 
open triangles the blue HB, open stars the extended blue HB tail, the green triangles 
the post HB, the open circle the blue stragglers candidates, and the asterisks are the blue 
hook candidates.}
\label{fig:box}
\end{figure*}

In Fig.~\ref{fig:box} the optical and UV band CMDs for both
clusters are shown. The most relevant, common properties of these two clusters are the
occurrence of an extended HB, and the tilt of the HB, as explained before.
However, the HB morphology in
the optical CMDs show also some differences: 
  the blue HB appears more
  populated and extended in NGC~6388 than in NGC~6441; 
  in the CMD of NGC~6441 there is a gap at $F555W\approx 18.5$; no such gap is
  visible in the HB of NGC~6388, which, instead seems to show a gap at
  $F555W\approx 19.8$; there is a large color spread at the hot end of the observed HB
sequence, of the order of $\sim0.5$ magnitudes in $F255W-F336W$,
larger than the photometric errors (less than 0.1 magnitudes).
%%%%%%%%% FIG 1 %%%%%%%%%%%%%
%%%%%%%%%%%%%%%%%%%%%%%%%%%%%%%

\subsection{Differential reddening}\label{sec:diffredd}
Both clusters are affected by a sizeable amount of
differential reddening (Piotto et al. 1997, Raimondo et al. 2002).
In order to estimate the size of this effect, I divided the Planetary
Camera field into 16 regions of $9\times$9 arcsec$^2$, and, for each
region, the corresponding CMD has been plotted (not shown here, see Fig.~4 of Busso et al.~2007).  
A slope in the HB consistent with that observed for the whole sample is evident also in those regions 
where the narrowness of the red giant branch suggests that no strong residual differential reddening is
present, forcing to conclude that the sloped HB is the consequence of an intrinsic
property of the HB stars in NGC~6388 and NGC~6441 and it is not caused by differential reddening. 

\subsection{Comparison with the models}
The observations were compared with Zero Age Horizontal Branch (ZAHB) models provided by Pietrinferni
et al.~(2006), supplemented by additional computations performed for
this specific project, with a metallicity Z=0.008 and He
contents Y=0.256 and an age of about 13Gyr. 
The reddening and distance
modulus values adopted are those for which ZAHB models with the canonical Y=0.256 best
fit the empirical distribution of the red HB clump in the (F336W-F439W,
F439W) CMD (see Fig. 2), i.e. $E(F439W-F555W)$=0.45 and $(m-M)_{F555W}$=16.65 for NGC~6388, while for NGC~6441 we adopt $E(F439W-F555W)$=0.48 and $(m-M)_{F555W}$=17.4.
%======== FIG 5 =============================================================== 
\begin{figure}
\begin{center}
\includegraphics[width=4.4cm,height=4.4cm]{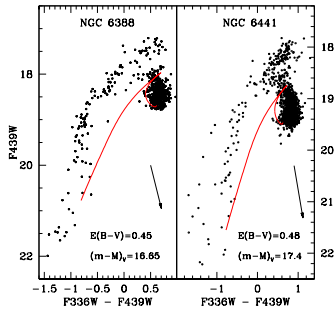}
\end{center}
\caption{(F336W-F439W, F439W) CMD of NGC~6388 and NGC~6441.  The
  reddening and distance modulus values determined by best fitting
 in this diagram  the red clump with the canonical ZAHB models (solid line) for $Y=0.256$ are labeled. These canonical ZAHB models are plotted by using a solid line. }
 \label{fig:fit_ub}
\end{figure}
%============================================================================
This arbitrary choice was adopted for the sake of consistency with previous works and because 
cool ZAHB models should be more reliable than hotter ZAHB models ($T_{eff}$ larger than about 10.000K), since they are not affected by diffusive processes.

\begin{figure*}
\begin{center}
\includegraphics[width=6cm,height=7cm]{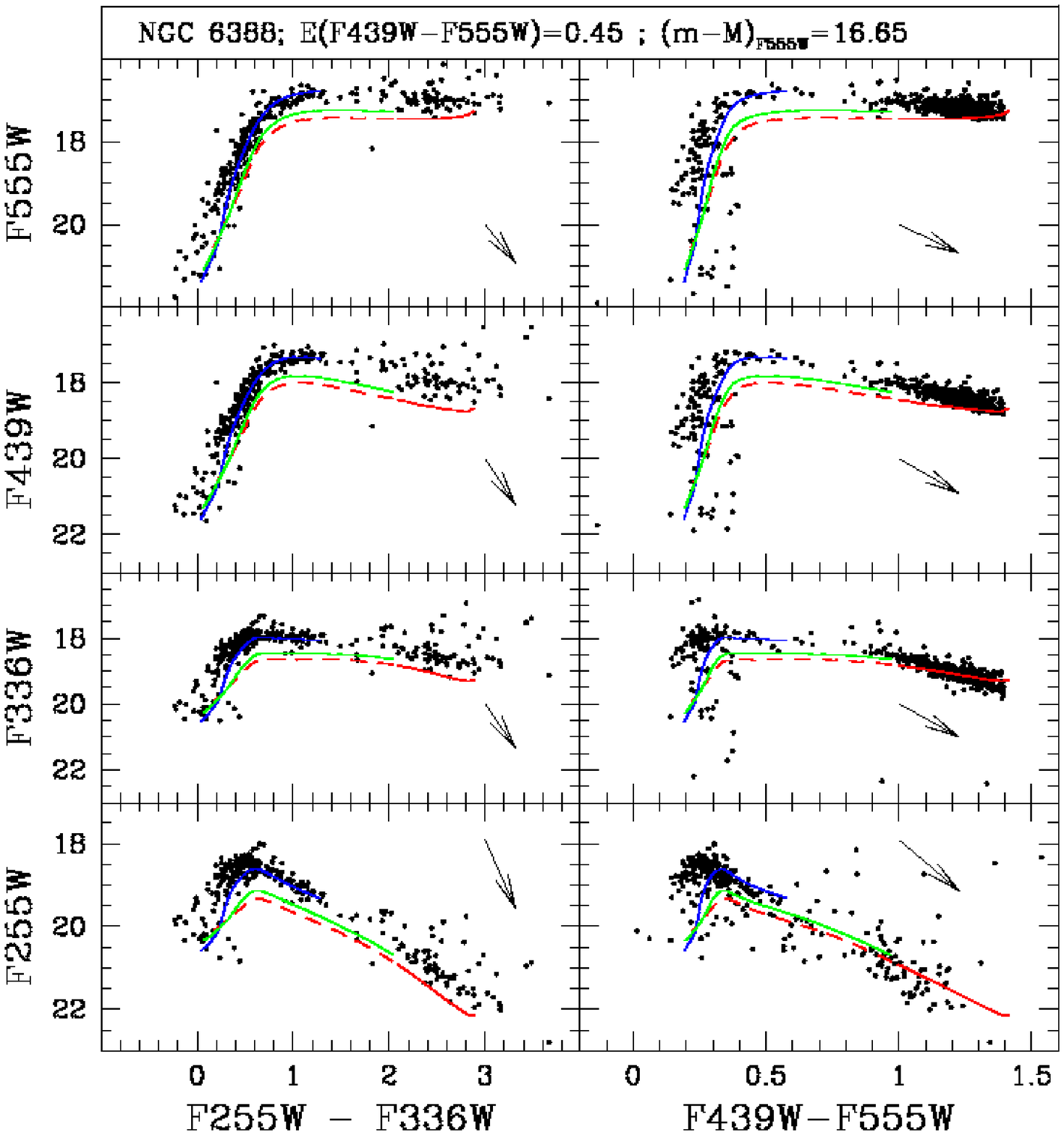}
\includegraphics[width=6cm,height=7cm]{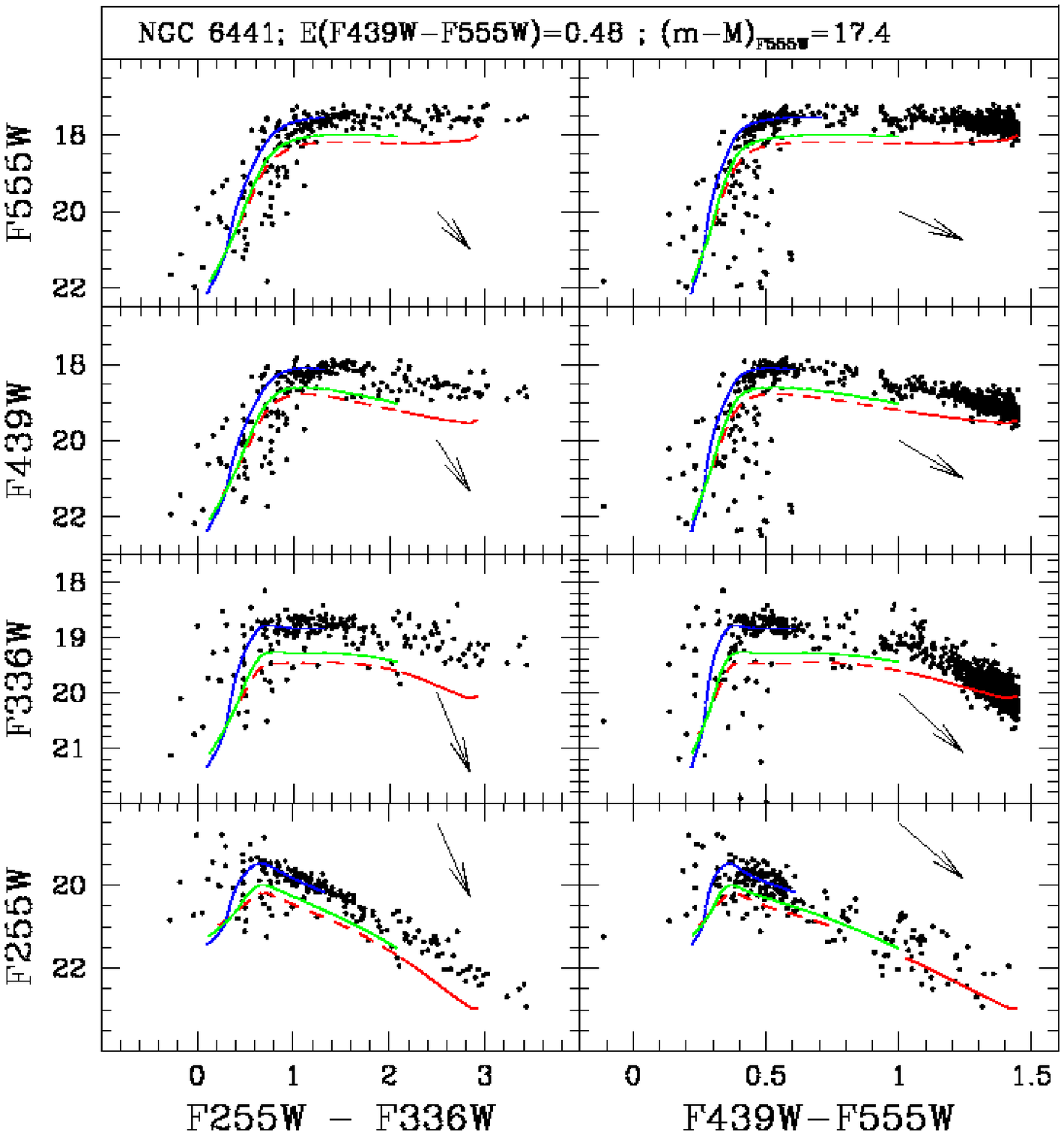}
\end{center}
\caption{\footnotesize 
Top panel: the HBs of NGC~6388 in different photometric planes. 
The observations are compared with  theoretical models for a canonical
Y=0.256 (lower red line), for Y=0.30 (middle green line), and for Y=0.40 (top blue line). 
The shifts applied to the stellar  models for the adopted distance modulus and mean reddening are labeled
in the figure, and they have been fixed according to the (arbitrary)
choice of imposing a best fit between models and the lower envelope of
the red HB portion in the (F336W-F439W, F439W) CMD (see Fig.~\ref{fig:fit_ub}). The arrows represent the
reddening vector in the different photometric planes.  Bottom panel: as in the top panel, but for NGC~6441.}
\label{fig:diff_cmd}
\end{figure*}

Figure~\ref{fig:diff_cmd} shows a comparison, in the various photometric planes,
between the ZAHB models and the HB sequences of NGC~6388 and NGC~6441. 
The ZAHB with canonical helium abundance Y=0.256 is plotted in the CMD with a red line.
The comparison between theory and observations clearly
shows that it is not possible to have an overall fit of the entire
HB: a good fit of the red HB do not allow a good fit for the blue part, and vice versa. 
The striking evidence is that this
is true for all combinations of colors and magnitudes, which raises
the suspicion that there must be a real mismatch between the canonical
ZAHB models and the observed HBs for both clusters.  
%%%%%%%%%%% FIG 6 %%%%%%%%%%%%%%%%%%
%%%%%%%%%%%%%%%%%%%%%%%%%%%%%%%%%%%%%%%%
Fig.~\ref{fig:diff_cmd} shows also that 
the models with canonical Y=0.256 are not able 
to reproduce the color of the hottest HB stars in the CMD. This empirical finding could be
interpreted as evidence that the effective temperature of these
stars is significantly higher than that of the hottest HB models.
This occurrence is evident for NGC~6388, but much less clear in
NGC~6441, which is not surprising, since we already noted that the latter
has a much less populated and less extended blue HB. 
The presence of a possible gap, at $F555W \sim 19.8$, in the case of NGC~6388, just beyond the location of the hottest stars in the canonical models suggests that the stars
observed beyond the gap could 
be blue hook stars, as  those ones already observed by
D'Cruz et al.~(2000) in $\omega$~Centauri (NGC~5139)
and by Brown et al.~(2001) in NGC~2808.

\subsection{A possible solution to the enigma}

An interesting scenario that could help to solve the puzzle of the
extended and tilted HBs of NGC~6388 and NGC~6441, has recently emerged
from some quite unexpected results obtained for the
Galactic GCs $\omega$~Cen and NGC~2808: both show multiple main sequence
(Bedin et al.~(2004), D'Antona et al. 2005, Piotto et al. 2007) 
which can be explained assuming that the various sequences are associated to stellar
populations characterized by different initial He
contents (D'Antona et al. 2005, Piotto et al. 2007). 

The fact that, with $\omega$~Cen and NGC~2808, 
NGC~6388 and NGC~6441 are between the most massive GCs
suggest  to associate the anomalously blue and
anomalously tilted HBs of NGC~6388 and NGC~6441 to a second generation
of stars, strongly He-enriched by pollution from massive and/or intermediate-mass stars of the
first star formation burst. 
A similar suggestion that the presence of a population with He
enhancement can explain the anomalous HB of NGC 6388 and NGC 6441 has
been recently made also by Caloi and D'Antona (2007).  
In order to verify this scenario
additional sets
of low-mass, He-burning models for a metallicity
Z=0.008 and inital He contents equal to Y=0.30 and 0.40, were calculated. 
Fig.~\ref{fig:diff_cmd} shows a comparison between the observed CMDs and
the theoretical models computed for the various He contents, adopting
the same reddenings and distance moduli. 
As already discussed, the models computed by assuming a canonical He content (Y=0.256) are
not able to reproduce the blue HB. On the contrary, one can easily see
that the ZAHB models corresponding to Y=0.40 are in good agreement
with the observed distribution of blue HB stars of NGC~6388 in all of
the CMDs (top of Fig.~\ref{fig:diff_cmd}).  The same kind of comparison, but for the case of
NGC~6441, is performed in the bottom panels of Fig.~\ref{fig:diff_cmd}. In this case, it is evident that the
Y=0.40 ZAHB is slightly brighter than the observed distribution of
blue HB stars.  From the comparison between empirical data and the ZAHBs
computed for various initial He contents, it appears that it is
possible to reproduce the brightness of the blue HB star population
by assuming a He content of the order of $Y\approx0.35$, i.e.
slightly lower than in NGC~6388. This fact is consistent with the
observational evidence that the blue HB is less extended in NGC~6441
than in NGC~6388.
The presence of two distinct stellar populations characterized by two
different initial He contents can help in explaining the brightness
difference between the red portion of the HB and the blue
component. However, in order to explain the tilted morphology of the
whole HB sequence one should also account for the presence of a spread
in the He content at the level of about $\Delta{Y}\approx0.05--
0.06$ (see also Moehler \& Sweigart 2006b,
Caloi \& D'Antona 2007). 

Summarizing, the main peculiarities of the HBs are their extension to very
hot temperatures, including (at least for NGC~6388) a number of blue
hook candidates, and the presence of a tilt in the HB. 
Reddening and differential reddening may contribute to create a sloped HB, but neither of these effects is
sufficient to explain the observed tilted HB.
Canonical models are not able to reproduce 
the peculiar HBs of NGC~6388 and NGC~6441, while 
the presence of a He-rich stellar component allows to explain the observed
blue HsB and the fact that they are brighter than the red HB clump. The presence of a He spread between the
\lq{canonical}\rq\ value and $Y\sim0.35 - 0.40$ allows to explain the observed
upward slope of the HB in both clusters.

\bibliographystyle{aa}

\end{document}